\documentclass[letter]{aa} % for the letters 
\usepackage{graphicx}
\usepackage{txfonts}

\begin{document}
   \title{A possible solution for the lack of EHB binaries 
          in globular clusters } 

   \author{Z. Han}
   \institute{National Astronomical Observatories / Yunnan Observatory,
              the Chinese Academy of Sciences, Kunming, 650011, China\\
              \email{zhanwenhan@hotmail.com}
             }

   \date{Received ; accepted }

% \abstract{}{}{}{}{} 
% 5 {} token are mandatory
 
  \abstract
  % context heading (optional)
  % {} leave it empty if necessary  
   {The binary fraction among extreme horizontal branch (EHB) stars
    in Galactic globular clusters (GCs) is an order of magnitude lower
    than the binary fraction among their 
    counterparts, field hot subdwarfs. 
    This casts serious doubt on their formation channels.}
  % aims heading (mandatory)
   {In this {\em Letter}, I explain the difference between 
    the field and the cluster EHB stars with the binary
    model of Han et al. (2002, 2003) for the formation of EHB stars.
   }
  % methods heading (mandatory)
   {With the binary population synthesis code of Han et al. 
   (2002, 2003), I follow the evolution of simple stellar populations
   resulting from single star bursts (note that 
   Han et al. 2002, 2003, adopted a constant star formation 
   rate over the last 15\,Gyr for the production of field EHB stars), 
   and obtain EHB stars at different stellar population ages. 
   }
  % results heading (mandatory)
   {I found that the binary fraction among EHB stars decreases 
    with the stellar population age.
    The fraction of EHB binaries with orbital periods $P<5\,{\rm d}$ 
    is $\sim 2.5\%$ for a stellar population of 10\,Gyr
    from the standard simulation set.
   }
  % conclusions heading (optional), leave it empty if necessary 
   {The binary model of Han et al. (2002, 2003) is able to explain
    the lack of EHB binaries in globular clusters.
    I also propose that the precise determination of the physical parameters
    of close EHB binaries in GCs can lead to the strictest constraint
    on common-envelope ejection efficiency.
   }


   \maketitle
%
%________________________________________________________________

\section{Introduction}
Extreme horizontal branch (EHB) stars (also known as hot subdwarfs)
are generally considered
to be helium-core-burning stars with extremely thin hydrogen envelopes
($<0.02M_\odot$)
(Heber \cite{heb86}; Saffer et al. \cite{saf94}).
They play an important role in many aspects of astrophysics,
e.g., stellar evolution, distance indicators, Galactic structure,
and the long-standing problem of far-ultraviolet excess 
in early-type galaxies 
(Kilkenny et al. \cite{kil97}; 
 Green, Schmidt \& Liebert \cite{gre86};
 Han, Podsiadlowski \& Lynas-Gray \cite{han07}).

Various scenarios have been proposed to explain the formation
of EHB stars  
(Mengel, Norris \& Gross \cite{men76}; 
 Webbink \cite{web84};
 Tutukov \& Yungelson \cite{tut90};
 Sweigart \cite{swe97}; 
 D'Cruz et al. \cite{dcr96}; 
 Lee \cite{lee94};
 Bressan, Chiosi \& Fagotto \cite{bre94};
 Yi, Demarque \& Oemler \cite{yi97}), 
and most of these scenarios are based on single star evolution
 (see, however, Mengel, Norris \& Gross \cite{men76})
or are meant for single EHB stars 
 (see Webbink \cite{web84}; Tutukov \& Yungelson \cite{tut90},
for the merger of two helium 
white dwarfs (WDs) to form a single EHB star).
Some of the models are also quite ad hoc 
(see Han, Podsiadlowski \& Lynas-Gray \cite{han07}, for a review).
Observationally, about two-thirds of field EHB stars
are in close binaries (Maxted et al. \cite{max01}), and
this presented a serious challenge to the theory for the formation
of EHB stars.  
Han et al. (\cite{han02,han03}) proposed an a priori binary model 
(hereafter HPMM model) for the formation of EHB binaries and
single EHB stars.
In the model, there are three types of formation channels for EHB stars,
involving stable Roche lobe
overflow (RLOF) for EHB binaries with long orbital periods,
common-envelope (CE) (Paczy\'nski \cite{pac76}) ejection for 
EHB binaries with short orbital periods, and the merger
of helium WDs to form single EHB stars.  The model
explains the main observational characteristics of field EHB stars: in
particular, their distributions in the orbital period-minimum
companion mass diagram, and in the effective temperature-surface
gravity diagram; their distributions of orbital period and mass
function; their binary fraction and the fraction of EHB
binaries with WD companions; their birth rates; and their
space density. The HPMM model is a significant advancement
and is widely used in the study of EHB stars 
(O'Tool, Heber \& Benjamin \cite{oto04}).

The majority of field EHB stars are in close binaries,
and HPMM model explained the binary fraction. 
Therefore, it was a great surprise that 
radial-velocity surveys revealed a remarkable lack of close 
binary systems in EHB stars in globular clusters (GCs)
(Moni Bidin et al. \cite{mon06}; Moni Bidin, Catelan \& Altmann \cite{mon08}).
This may imply that the formation channels for cluster EHB stars
are different from those for field EHB stars. 
However, by checking the channels in the HPMM model and the model results,
Moni Bidin, Catelan \& Altmann (\cite{mon08}) speculated that
a binary fraction-age relation may be
responsible for the lack of EHB binaries.
Does the relation really exist ? If it exists, is it able to explain
the lack of EHB binaries in GCs ?

In this {\em Letter}, I carry out detailed Monte Carlo simulations
for the formation of EHB stars in GCs with the binary population synthesis 
code used for the HPMM model, and obtain the binary fraction-age
relation, which can explain the lack of EHB binaries in GCs.

\section{The model}
%_____________________________________________________________
%
\begin{table}
\caption{Parameters for simulation sets}
\label{table1}
\centering    
\begin{tabular}{c c c c c}
\hline\hline              
Set & $n(q')$ & $q_{\rm crit}$ & $\alpha_{\rm CE}$ & $\alpha_{\rm th}$ \\
\hline 
   1 & constant & 1.5 & 0.75 & 0.75 \\
   2 & constant & 1.5 & 0.50  & 0.50 \\
   3 & constant & 1.5 & 1.0 & 1.0 \\
   4 & constant & 1.2 & 0.75 & 0.75 \\
   5 & uncorrelated & 1.5 & 0.75 & 0.75 \\ 
\hline                                   %inserts single line
\end{tabular}
\end{table}

In the HPMM model, EHB stars are mainly produced from three types of channels:
stable RLOF, CE ejection, and the merger of helium WD pairs.
For a 'primordial' binary system, the primary (initially more massive
component) evolves and may fill its Roche lobe 
when it has a helium core. Given the mass ratio $q$ of the primary to
the secondary less than a critical value, the mass 
transfer is dynamically stable and leads to the formation
of a wide EHB binary with a main-sequence (MS) companion if the
helium core is ignited later
({\em 1st stable RLOF channel}). Given the mass ratio
over the critical value, a CE is formed, and the CE 
ejection produces a close EHB binary with a MS companion
if the helium core is ignited later ({\em 1st CE ejection channel}).
Some of the 'primordial' binary systems may evolve to WD binary systems
after the first mass transfer phase. In this case, the secondary 
continues to evolve and may fill its Roche lobe when it has a helium core.
If the mass transfer is stable,
a wide EHB binary with a WD companion is resulted given that the helium core
is ignited later ({\em 2nd stable RLOF channel}).
The mass transfer is very likely to be dynamically unstable,
and this results in the formation of a CE, and the CE ejection
produces a close EHB binary with a WD companion if the helium core is
ignited later ({\em 2nd CE ejection channel}). 
For a WD binary system,
if the WD is a helium type and the secondary is on the first giant branch
(i.e., containing a helium core) when it fills its Roche lobe,
the resultant CE contains helium WD pairs.
The ejection of the CE leads to the formation of close helium WD pairs,
and the close helium WD pairs may merge due to angular momentum loss of
gravitational wave radiation to form single EHB stars
({\em the merger channel}).

Binaries with primary's initial mass, 
$M_{\rm 1i}\sim 0.95\,-\,7M_\odot$ and initial orbital period 
$P_{\rm i}\sim 1\, -\, 1000\, {\rm d}$, may produce EHB stars
via the channels described above. 
For the 1st stable RLOF channel, the primary fills its Roche lobe
on the Hertzsprung gap or the first giant branch, and the mass transfer
strips the primary of its envelope and leaves a naked helium core. 
This leads to a wide EHB+MS binary, if helium is ignited. For a stellar
population of age $t<1\,{\rm Gyr}$ ($M_{\rm 1i}\ga 2M_\odot$), a binary
with $P_{\rm i}\sim 1\,-\,100\,{\rm d}$ can evolve to a wide EHB+MS
binary via stable RLOF on the Hertzsprung gap or 
the first giant branch, and a larger EHB mass corresponds to a small
$t$ (a large $M_{\rm 1i}$). For $t>1\,{\rm Gyr}$ ($M_{\rm 1i}\la 2M_\odot$),
only RLOF near the tip of the first giant branch (i.e., with a narrow
range of $P_{\rm i}$, typically $\Delta \log (P_{\rm i}/{\rm d})\sim 0.5$) 
can lead to
EHB binaries, as the naked helium core is not ignited if
the RLOF is not close to the tip of the first giant branch
(see Table 4 of Han et al. \cite{han02}). The corresponding
orbital period $P_{\rm i}$ is from $\sim 10\,{\rm d}$ to $\sim 1000\,{\rm d}$
for a stellar population with age $t$ of 1\,Gyr to 15\,Gyr.
The orbital period and the EHB mass of the resultant EHB binary
would be larger for a large $t$.  
For the 1st CE channel to form an EHB binary,
the primary of a binary system needs to fill its Roche lobe
while it has a helium core. However, the resultant CE cannot be
ejected to form a close EHB binary due to a tight envelope if 
$t<1\,{\rm Gyr}$ ($M_{\rm 1i}\ga 2M_\odot$).
For $t>1\,{\rm Gyr}$, the primary of a binary system needs
to fill its Roche lobe very close to the tip of the first
giant branch (i.e., with a typical
range of initial orbital period $\Delta \log (P_{\rm i}/{\rm d})\sim 0.1$)
(see Fig. 1 of Han et al. \cite{han02}), otherwise the helium
core cannot be ignited after the CE ejection. 
A large $t$ (a small $M_{\rm 1i}$) corresponds to a larger $P_{\rm i}$
($P_{\rm i} \sim 100\,-\, 1000\,{\rm d}$ for $t\sim 1\,-\,15\,{\rm Gyr}$)
and a larger orbital period $P$ of the resultant EHB binary.
This channel produces a close EHB binary with 
a typical EHB mass of $\sim 0.46M_\odot$.
No EHB star can form from the 2nd stable RLOF channel in the HPMM
model, as the mass ratio of the secondary (with an appropriate helium core)
to its WD primary is too large and the RLOF is not stable.  
For the second CE channel to form an EHB star,
the primary of a binary system first needs to experience a stable RLOF
(with $P_{\rm i}\sim 10\,-\,1000\,{\rm d}$)
to form a WD binary, and the secondary of the WD binary 
needs to fill its Roche lobe when its helium core mass is in an appropriate
range (see Fig. 1 of Han et al. \cite{han02}).
The ejection of the resultant CE leads to a close WD+EHB binary.
As the WD can spiral in deeper in the envelope during the CE ejection, 
the WD+EHB can have a much shorter orbital period 
(as short as $\sim 0.02\,{\rm d}$ for a small $t$) than
the MS+EHB binary from the first CE ejection channel.
For a large $t$, the WD spirals in the envelope of a less massive
secondary (the envelope is more loosely bound), 
and the resultant WD+EHB binary is wider.
For the merger channel,
a binary system first needs to experience a stable RLOF
(with $P_{\rm i}\sim 4\,-\,250\,{\rm d}$ and 
$M_{\rm 1i}\sim 0.95\,-\,2\,M_\odot$) to produce a
helium WD binary, and the binary experiences a CE evolution
to form a helium WD pair. The pair may coalesce to form
a single EHB star due to angular momentum loss of
gravitational wave radiation.
The EHB star produced from this channel has a wider mass range
($0.4\,-\,0.8M_\odot$) than from other channels. The mass range
is smaller for a small $t$ ($0.56\,-\,0.64M_\odot$ for $t=2\,{\rm Gyr}$)
as only very close helium WD pairs
have time enough to merge, but the range becomes larger with a large $t$ 
($\sim 0.4\,-\,0.8M_\odot$ for $t=15\,{\rm Gyr}$).

In the process of CE ejection, the orbital energy released by the
orbital decay of the embedded binary is used to 
overcome the binding energy of the CE. As is usual, I
defined two parameters: the
CE ejection efficiency $\alpha_{\rm CE}$, i.e.\ the fraction of the
released orbital energy used to overcome the binding energy; 
and $\alpha_{\rm th}$, which defines the fraction of the thermal
energy contributing to the binding energy of the CE.
As in the HPMM model, I adopted
$q_{\rm crit}=1.5$
(for stable RLOF on the first giant branch), and
$\alpha_{\rm CE}=\alpha_{\rm th}=0.75$
as the best choices, and varied them to see their effects.

To obtain the distributions of properties of EHB stars at different 
ages, I have performed detailed Monte Carlo simulations with 
the binary population synthesis code developed for the HPMM model.  In the
simulation, I followed the evolution of 10 million sample binaries
according to grids of stellar models of solar metallicity and the
evolution channels leading to EHB stars.
I adopted the following input for the simulations
(see Han, Podsiadlowski \& Eggleton \cite{han95}): 
\begin{description}
\item (1) A single star burst is adopted (note that the
    star-formation rate is taken to be constant over the last 15\,Gyr 
    in the HPMM model).

\item (2) The initial mass function of Miller \& Scalo (\cite{mil79}) 
    is adopted. 

\item (3) For the initial mass ratio distribution $n(q')$ 
    (where $q'$ is the ratio
    of secondary to primary), I adopted (a) a constant
    mass ratio distribution, or alternatively (b) a mass ratio distribution
    where the masses are uncorrelated and 
    drawn independently from a Miller \& Scalo (\cite{mil79}) initial
    mass function.

\item (4) I take the distribution of
    separations to be constant in $\log a$ for wide binaries, where $a$ is
    the orbital separation.  The adopted distribution implies that $\sim
    50$\,\% of stellar systems are binary systems with orbital periods
    less than 100 yr.

\end{description}

Similar to the HPMM model, I have carried out 5 sets of Monte Carlo
simulations altogether for Population I by varying the model parameters
over a reasonable range (see Table~\ref{table1}). As in the HPMM model,
the standard simulation set (set 1) has a constant initial mass ratio 
distribution, $q_{\rm crit}=1.5$, and $\alpha_{\rm CE}=\alpha_{\rm th}=0.75$. 

\section{Results}

\begin{figure}
  \centering
  \includegraphics[angle=-90,width=8cm]{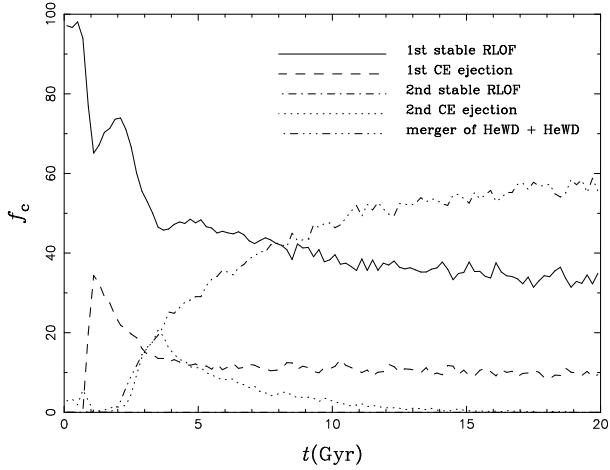}
  \caption{
  The evolution of the fraction $f_{\rm c}$ 
  (in percentage) of EHB stars originated 
  from each channel for the standard simulation set (set 1), i.e.,
  the 1st stable RLOF channel (wide EHB+MS binaries);
  the 1st CE ejection channel (close EHB+MS binaries);
  the 2nd stable RLOF channel (wide EHB+WD binaries);  
  the 2nd CE ejection channel (close EHB+WD binaries); 
  and the merger channel (single EHB stars).
  No selection effect has been applied to the lines.
  Note that there are no EHB binaries from the 2nd stable RLOF channel.
  }
  \label{channel}
\end{figure}

\begin{figure}
  \centering
  \includegraphics[width=8cm]{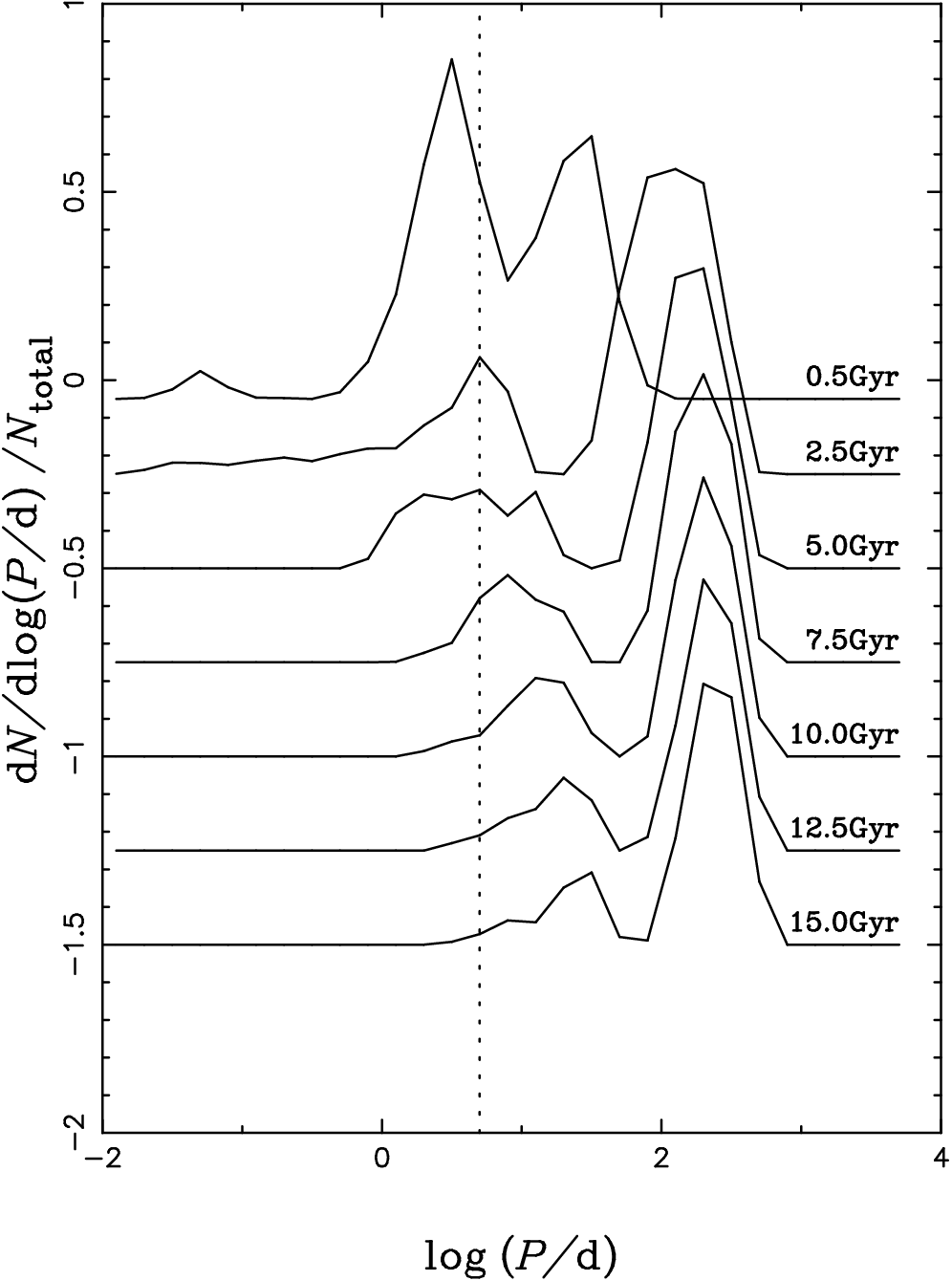}
  \caption{The evolution of the distributions of 
  orbital periods with stellar population age $t$ 
  for the standard simulation set (set 1).
  $P$ is orbital period, and
  $N_{\rm total}$ is the total number of EHB stars, including both 
  binaries and singles. The stellar population ages are denoted and 
  each distribution
  has been applied an offset of $-0.1\times (t/{\rm Gyr})$, 
  where $t$ is the age.
  The distributions are without any selection effect. If we apply the GK
  selection effect, the far-right peaks, which are mainly EHB binaries with
  more massive MS companions from the 1st stable RLOF channel, 
  disappear. The vertical dotted line denotes $P=5\,{\rm d}$.
  }
  \label{period}
\end{figure}

\begin{figure}
  \centering
  \includegraphics[angle=-90,width=8cm]{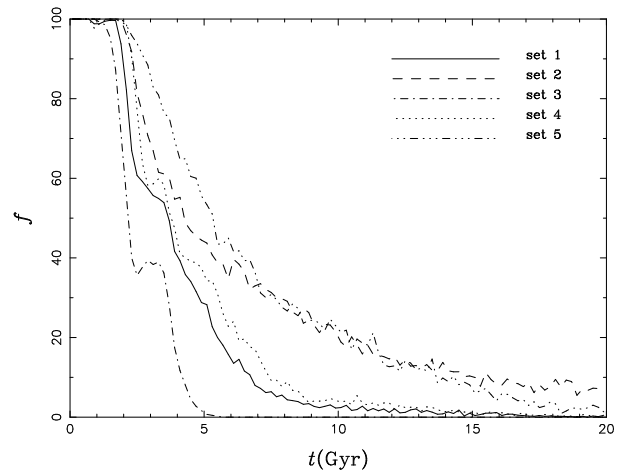}
  \caption{The evolution of the fraction $f$ (in percentage) 
  of close EHB binaries 
  (with orbital periods $P<5\,{\rm d}$) among all EHB stars,
  including both binaries and singles.
  Note that the GK selection effect is applied. If the selection effect
  is not considered, $f$ would be smaller due to the fact that the GK 
  selection effect 
  excludes wide EHB+MS binaries from the first stable RLOF channel.
  }
  \label{close}
\end{figure}

The simulations give EHB stars at various stellar population ages.
In order to see the importance of individual evolution channels
leading to the formation of EHB stars, I plotted in 
Fig.~\ref{channel} the fractions of EHB stars from
different channels 
(the 1st stable RLOF channel for wide EHB+MS binaries; 
the 1st CE ejection channel for close EHB+MS binaries; 
the 2nd stable RLOF channel for wide EHB+WD binaries;
the 2nd CE ejection channel for close EHB+WD binaries; 
and the merger channel for single EHB stars) at stellar population
age $t$ for the 
standard simulation set (set 1).

Figure~\ref{period} shows 
the evolution of the distribution of orbital periods of EHB binaries 
with stellar population age $t$ for the standard simulation set (set 1).
The EHB binaries with orbital periods $P<5\,{\rm d}$ can be
detected observationally in a GC 
(Moni Bidin, Catelan \& Altmann \cite{mon08}), and  I, therefore,
showed, in Fig.~\ref{close}, 
the fractions of the binaries among all the EHB stars, including
both binaries and singles, at stellar population age $t$.
In order to see how the fractions can
be affected by model parameters, Fig.~\ref{close} also displays other
simulation sets with various model parameters.  
In the figure, I have applied the GK selection effect,
which is the selection against EHB stars with 
companions of spectral type G and K 
(usually MS stars), and 
is the most important selection effect in observations
of EHB stars\footnote{
We applied the GK selection effect in the following way.
If an EHB binary has a MS 
companion and the effective temperature
of the companion is above 4000\,K or the companion is brighter than
the EHB star, the system is excluded. See the HPMM model for details.}.

\section{Discussion}

Figure~\ref{channel} shows the fraction of EHB stars originated 
from each channel. We see that the 1st stable RLOF is dominant for
$t<8\,{\rm Gyr}$. The resultant wide MS+EHB binaries, however, may
not be observed due to the GK selection effect as the MS secondaries
are too massive and bright. The 1st CE ejection channel
and the 2nd CE ejection channel have a contribution peak
at $t\sim 1.2\,{\rm Gyr}$ and $t\sim 3.5\, {\rm Gyr}$, respectively.
The merger channel starts to contribute noticeably at $t\sim 2\,{\rm Gyr}$
and dominates at $t\ga 8\,{\rm Gyr}$, or at $t\ga 3.5\,{\rm Gyr}$ if
the GK selection is considered.
  
Figure~\ref{period} shows the distributions of orbital periods
of EHB binaries at various ages of a stellar population.
For $t=0.5\,{\rm Gyr}$, the distribution has 3 peaks.
The left peak (the small one at $\log (P/{\rm d}) \sim -1.3$)
is from the 2nd CE ejection channel, where a
WD has spiralled in deeply into a tightly bound envelope of its companion
(with $M_{\rm 2i}>2M_\odot$) to form a close WD+EHB binary.
Both the middle peak and the right peak are from the 1st stable
RLOF channel with $M_{\rm 1i}>2M_\odot$, 
while the middle peak is from stable RLOF on the Hertzsprung
gap (with a mass ratio of $q_{\rm i}\sim 1\,-\,3$) and
the right peak from stable RLOF on the first giant branch (with a
mass ratio of $q_{\rm i}\sim 1\,-\,1.5$).
For $t=2.5\,{\rm Gyr}$, the distribution has two peaks.
The left peak is from the 1st CE ejection channel
and the part to the left of the peak is from the 2nd CE ejection channel.
The right peak is from the 1st stable RLOF channel,
and the orbital period corresponding to the right peak is significantly larger 
than that at $t=0.5\,{\rm Gyr}$, which is due to the fact
that stellar radius at the tip of the first giant branch is 
significantly larger
for $M_{\rm 1i}< 2M_\odot$ than for $M_{\rm 1i}>2M_\odot$. 
For $t=5\,{\rm Gyr}$ or larger $t$, the 2nd CE ejection
channel produces WD+EHB binaries with orbital periods   
similar to that of MS+EHB binaries from the 1st CE ejection channel,
as the mass donors (the primary for the 1st CE ejection channel, or
the secondary for the 2nd CE ejection channel) are less massive 
than $2M_\odot$, and the envelopes are similarly loosely bound near the tip
of the first giant branch. 
For a large $t$, the radius (and the corresponding $P_{\rm i}$)
at the tip of the first giant branch is
bigger for the donor, and the envelope is also more loosely bound, and
therefore the left peak moves toward a longer orbital period for a large $t$.

The EHB binaries with orbital periods $P<5\,{\rm d}$ are from
the 1st CE ejection channel or the 2nd CE ejection channel.
Figure~\ref{close} shows the evolution of the fraction $f$
of close EHB binaries (with $P<5\,{\rm d}$) with stellar population age $t$.
We see that $f$ decreases sharply with $t$, which is largely
due to the increasing importance of the merger channel with $t$
(see Fig.~\ref{channel}).
For the standard simulation set (set 1),
$f$ is $\sim 2.5\%$ for $t=10\,{\rm Gyr}$.
In order to see the effects of model parameters,
I performed 5 sets of simulations altogether.
Set 2 adopted smaller common envelope ejection efficiency
and the orbital periods of close EHB binaries are, therefore, smaller,
leading to bigger $f$ for $t>1\,{\rm Gyr}$.
Set 3, on the other hand, adopted bigger efficiency and,
therefore, $f$ is smaller. 
Set 4 adopted smaller $q_{\rm c}$ for stable RLOF
on the first giant branch, and this makes CE more likely, and
leads to more EHB binaries from CE ejection channels, which
results in bigger $f$. 
Set 5 adopted an uncorrelated initial mass ratio distribution,
which means a large initial mass ratio $q_{\rm i}$ is more likely
and as a consequence CE is more likely experienced and, therefore,
$f$ is larger.
Note that the parameters adopted for the standard simulation set
(set 1) produce good matches between observation
and theory for field EHB stars (see HPMM for details).

There is a very high fraction of close EHB binaries in the field 
($\sim 2/3$, see Maxted et al. \cite{max01}), but there is a lack
of close EHB binaries in GCs. 
Moni Bidin et al. (\cite{mon06}) and
Moni Bidin, Catelan \& Altmann (\cite{mon08}) found that
the binary fraction $f$ (for $P<5\,{\rm d}$) is very small in NGC 6752,
with a most likely value of $f=4\%$ and an upper limit of 
$f=16\%$ at the 95\% confidence level.
They speculated that the sharp contrast between field EHB stars
and GC EHB stars may be due to a possible $f$-$t$ 
relation\footnote
{Napiwotzski et al. (\cite{nap04}) invoked
a possible $f$-$t$ or $f$-$\rm [Fe/H]$ relation to explain
the binary fraction difference between their survey for EHB stars 
and previous surveys for field EHB stars.
Their sample contains more thick disk or halo stars. 
}.
They analysed the HPMM model and Fig.~7 of Han, Podsiadlowski
and Lynas-Gray (\cite{han07}) and found that such a relation
is possible\footnote{
Moni Bidin, Catelan \& Altmann (\cite{mon08}) 
also argued that dynamical effects may not 
contribute to the formation
of EHB stars in GCs, as there is a lack of radial gradients
among EHB stars in NGC 2808 (Bedin et al. \cite{bed00})
and $\omega$ Centauri (D'Cruz et al. \cite{dcr00}).}.
This {\em Letter} showed that the $f$-$t$ relation
clearly exists, in support
of the speculation of Moni Bidin, Catelan \& Altmann (\cite{mon08}).
For field EHB stars, their progenitors have a wide spectrum
of masses due to the continuous star formation rate during the past,
and young (massive) progenitors are more likely to produce closer 
EHB binaries, 
and therefore the binary fraction is much higher (see the HPMM model).

Gratton et al. (\cite{gra03,gra05}) 
derived the average metallicity and the age of NGC 6752 to be 
${\rm [Fe/H]}=-1.48\pm 0.07$ and $13.8\pm1.1\,{\rm Gyr}$, respectively.
The standard simulation set (set 1) gives the fraction of close 
EHB binaries with $P<5\,{\rm d}$ to be $f\sim 1.3\%$ 
at $t=13.8\,{\rm Gyr}$, close to
the fraction of $4\%$ observed by 
Moni Bidin, Catelan \& Altmann (\cite{mon08})
if we consider that
$f$ is very sensitive to the model parameters, e.g. CE ejection efficiency, 
in each simulation set (see Fig.~\ref{close}).
Note the parameters adopted for the standard simulation set
give a high binary frequency among field EHB stars
($\sim 55\% $ after selection effects are taken, see Table 2 of
Han et al. \cite{han03}), consistent with observations of field EHB stars.  
Note also that our model is for solar metallicity.
For low metallicity, a star has a more 
tightly-bound envelope. Therefore, the EHB binaries from CE ejections
have shorter orbital periods, and the helium WD pairs would be closer
and should be more likely to merge to form single EHB stars.
However, the production of EHB stars are not expected to be affected much
by metallicity in the HPMM binary model.

The CE ejection efficiency strongly affects the distribution of close 
EHB binaries (see Fig.~\ref{close}). 
If we can obtain the parameters, i.e., the orbital period, the
mass of the EHB star, and the mass of the companion, of a close EHB binary in
a GC of known age and metallicity, we may constrain the CE ejection 
efficiency. From the age and metallicity of a GC, 
we can infer the mass of the progenitor of the EHB star. From the mass
of the EHB star and the mass of the companion, 
we would get the location of the CE on the evolutionary track of the 
progenitor, and we would then know the binding energy of the CE and the
separation before the CE. The separation after the CE can be easily
calculated from the binary parameters observed.
Then it would be straight forward to derive the CE ejection efficiency.
I, therefore, propose here that precise determination of the physical
parameters of close EHB binaries in GCs would possibly give the 
strictest constraint on CE ejection efficiency.
However, such an approach seems to be too remote. A more realistic
observational test of the results in this {\em Letter}
would be searching for EHB binaries with orbtial
periods of $\sim 10\,-\,20\, {\rm d}$ in a GC (see Fig.~\ref{period}).

\begin{acknowledgements}
I thank an anonymous referee for his/her generous comments which
helped to improve the paper, and 
Ph. Podsiadlowski for stimulating discussions.
This work was in part supported by the Natural Science Foundation of China 
under Grant Nos 10433030, 10521001 and 2007CB815406.
\end{acknowledgements}


\begin{thebibliography}{}
\bibitem[2000]{bed00}
 Bedin, L.R., Piotto, G., Zoccali, M., et al. 2000, A\&A, 363, 159
\bibitem[1994]{bre94}
 Bressan, A., Chiosi, C., \& Fagotto, F. 1994, ApJS, 94, 63
\bibitem[1996]{dcr96}
 D'Cruz, N.L., Dorman, B., Rood, R.T., \& O'Connell, R.W. 1996,
 ApJ, 466, 359
\bibitem[2000]{dcr00}
 D'Cruz, N.L., O'Connell, R.W., Rood, R.T., et al. 2000, ApJ, 530, 352
\bibitem[2003]{gra03}
 Gratton, R.G., Bragaglia, A., Carretta, E., et al. 2003, A\&A, 408, 529
\bibitem[2003]{gra05}
 Gratton, R.G., Bragaglia, A., Carretta, E., et al. 2005, A\&A, 440, 901
\bibitem[1986]{gre86}
 Green, R.F., Schmidt, M., \& Liebert, J. 1986, ApJS, 61, 305
\bibitem[1995]{han95}
 Han, Z., Podsiadlowski, Ph., \& Eggleton, P.P. 1995, MNRAS, 272, 800
\bibitem[2002]{han02}
 Han, Z., Podsiadlowski, Ph., Maxted, P.F.L., Marsh, T.R., \& Ivanova, N.
 2002, MNRAS, 336, 449
\bibitem[2003]{han03}
 Han, Z., Podsiadlowski, Ph., Maxted, P.F.L., \& Marsh, T.R.
 2003, MNRAS, 341, 669
\bibitem[2007]{han07}
 Han, Z., Podsiadlowski, Ph., \& Lynas-Gray, A.E.
 2007, MNRAS, 380, 1098
\bibitem[1986]{heb86}
 Heber, U. 1986, A\&A, 155, 33
\bibitem[1997]{kil97}
 Kilkenny, D., Koen, C., O'Donoghue, D., \& Stobie, R.S. 1997,
 MNRAS, 285, 640
\bibitem[1994]{lee94}
 Lee, Y.W. 1994, ApJ, 430, L113
\bibitem[2001]{max01}
 Maxted, P.F.L., Heber, U., Marsh, T.R., \& North, R.C. 2001, MNRAS, 326, 1391
\bibitem[1976]{men76}
 Mengel, J.G., Norris, J., \& Gross, P.G. 1976, ApJ, 204, 488
\bibitem[1979]{mil79}
 Miller, G.E. \&  Scalo, J.M. 1979, ApJS, 41, 513
\bibitem[2008]{mon08}
 Moni Bidin, C., Catelan, M., \& Altmann, M. 2008, A\&A, 480, L1
\bibitem[2006]{mon06}
 Moni Bidin, C., Moehler, S., Piotto, G., et al. 2006, A\&A, 451, 499
\bibitem[2004]{nap04}
 Napiwotzki, R. Carl, C.A., Lisker, T., et al. 2004, Ap\&SS, 291, 321
\bibitem[2004]{oto04}
 O'Tool, S.J., Heber, U., \& Benjamin, R.A. 2004, A\&A, 422, 1053 
\bibitem[1976]{pac76} 
 Paczy\'nski, B. 1976, Common Envelope Binaries, in Structure and Evolution of Close Binaries,
 ed. P.P. Eggleton, S. Mitton, \& J. Whelan (Kluwer, Dordrecht), 75
\bibitem[1994]{saf94}
 Saffer, R.A., Bergeron, P., Koester, D., \& Liebert, J. 1994, ApJ, 432, 351
\bibitem[1997]{swe97}
 Sweigart, A.V. 1997, ApJ, 474, L23
\bibitem[1990]{tut90} 
 Tutukov, A.V., \& Yungelson, L.R. 1990, A.Zh., 67, 109
\bibitem[1984]{web84}
 Webbink, R.F. 1984, ApJ, 277, 355
\bibitem[1997]{yi97}
 Yi, S.K., Demarque, P., \& Oemler, A. Jr. 1997, ApJ, 486, 201

\end{thebibliography}
\end{document}